\documentclass[psfig,10pt]{article}

\usepackage{inconsolata}
\usepackage[english]{babel}
\usepackage[utf8x]{inputenc}
\usepackage{amsmath}
\usepackage{graphicx}
\usepackage{listings}
\usepackage{color}
\usepackage{upquote}
\usepackage{underscore}

\definecolor{grey}{rgb}{0.5,0.5,0.5}

\title{Making Systems More Robust with Flexible RPC Lookup}
\author{Russell Power}
\date{}

\begin{document}
\maketitle

\begin{abstract} 
  
Modern distributed systems use \emph{names} everywhere.  Lockservices such as
Chubby~\cite{chubby} and ZooKeeper~\cite{zookeeper} provide an effective
mechanism for mapping from application names to server instances, but proper
usage of them requires a large amount of error-prone boiler-plate code.

Application programmers often try to write wrappers to abstract away this logic,
but it turns out there is a more general and easier way of handling the issue.  
We show that by extending the existing name resolution capabilities of 
RPC libraries, we can remove the need for such annoying boiler-plate code while
at the same time making our services more robust.  

\end{abstract}

\section{Introduction}

Lockservers are gaining traction as an effective way to handle the thorny issue
of ownership in distributed systems.  Lock-services typically manage a
directory of \emph{leases}, each of which maps a name to a server that owns the
lease.  The advantage of using a lock-service is obvious - we only have to
implement a distributed consensus algorithm in one place, and we can use it for
all of our distributed systems.

Far too frequently, however, we these services used in the following
way (in this example, looking up the owner of a tablet in BigTable~\cite{bigtable}):

\begin{lstlisting}[caption=Simple BigTable lookup.]
tablet_name = tablet_for_key("key")
server_name = lockserver.lookup(tablet_name)
server = rpc.connect(server_name)
server.put("key", "value")
\end{lstlisting}

This code is innocuous, but behaves poorly in the event of server failures.
Assuming that the server side of this operation correctly rejects the 
operation, we're still going to have a missing put.  Hopefully the programmer
realizes this (or their RPC implementation throws exceptions).  If not, 
we lose the put operation.  If the programmer does realize the put can fail,
we end up wrapping our logic in a cumbersome retry loop which does not make our
code any easier to read:

\begin{lstlisting}[caption=Fixed BigTable lookup.]
timeout = 1
while 1:
  tablet_name = tablet_for_key("key")
  server_name = lockserver.lookup(tablet_name)
  server = rpc.connect(server_name)
  server.put("key", "value", timeout=timeout)
  timeout = min(timeout * 2, 60)
  if result.success(): break
\end{lstlisting}

A cursory examination of code that uses these lock services finds that these
retry-loops occur very frequently.  (Indeed, even the ZooKeeper codebase
contains wrapper functions to try to simplify these issues when talking to the
leasing service itself). 

It turns out there is a very straightforward and robust way to encapsulate this
retry pattern: extend our RPC system to use lease keys as names. Not only does
this remove the need for retry logic, lookups are now handled implicitly,
simplifying our code even further.  What's more, we even get a good guess at
our what sort of timeout value to use for free (the remaining lease time).  Most
RPC libraries already have this sort of code for dealing with more mundane 
re-connection issues already.  Any RPC implementation worth it's salt already 
has built-in support for one particular instance of this behavior, in the form 
of hostname (DNS) resolution.  (Some even support looking up lease names,
but sadly they are typically just used as an alternate form of domain name 
lookup).

To see what this looks like -- our RPC library client gets extended to look up 
servers directly via the lease name:

\begin{lstlisting}[caption=Resolution integrated into RPC library.]
class RPCClient:
  def call(lease_name, request):
    while 1:
      target, lease_time = lockserver.lookup(lease_name)
      socket = connect(target)
      result = socket.send(request.str(), timeout=lease_time)
      if result.timed_out(): continue
      else:
        # success or application level error
        return result
\end{lstlisting}

Now in our application code we can use the leases directly, removing the lookup
step and making our code more robust at the same time:

\begin{lstlisting}[caption={Simple BigTable lookup, take 2.}]
tablet_name = tablet_for_key("key")
tablet = rpc.connect(tablet_name)
tablet.put("key", "value")
\end{lstlisting}

Note that it's now clearer that our operation is on a particular \emph{tablet}.
But of course, if you're paying close attention (and aware of the details of how
BigTable works), you'd notice that we might be talking to the wrong tablet.  If
the original tablet split between the time we got our name and when we did the
put, then our client won't have anyone to talk to -- a very sad situation
indeed.  We might be tempted to wrap this into a retry loop like we did above,
but then we'd have the same ugly, error-prone logic again:

\begin{lstlisting}[caption={Simple BigTable lookup, take 3.}]
while 1:
  tablet_name = tablet_for_key("key")
  tablet = rpc.connect(tablet_name)
  result = tablet.put("key", "value", timeout=XXX)
  if result:
    ...
  
\end{lstlisting}

We can do better though, and remove the yucky direct dependency on a
lockservice we added to our RPC at the same time -- we just to extend the
concept of "lookup" for our RPC library.  Traditionally, RPC libraries have
hard-coded paths for handling things like domain-name resolution.  But modern
distributed applications have more flexible names -- we should have more
flexible naming.  This is easily achieved by just specifying how to do lookup
as a separate function.  Note that our lookup function itself might now make
RPC calls of it's own.

\begin{lstlisting}[caption=Resolution integrated into RPC library.]
class RPCClient:
  def call(request, lookup_fn):
    while 1:
      target, timeout_guess = lookup_fn(request)
      socket = connect(target)
      result = socket.send(request.str(), timeout=timeout_guess)
      if result.timed_out(): continue
      else:
        # success or application level error
        return result
\end{lstlisting}

And our client code becomes:

\begin{lstlisting}[caption={Simple BigTable lookup, take 4.}]
def lookup_tablet(req, next):
  tablet_name = tablet_for_key(req.key)
  return next(tablet_name)

# try applying each lookup function in order until we get 
# a match
session = rpc.session(lookup=
  [ lookup_tablet, lookup_lease, lookup_dns ])

result = session.put("key", "value")
\end{lstlisting}

A simple local form of this request-based lookup is frequently seen in
message-queue based systems (typically for sharding a request to a number of
workers).  Note that we've also reduced the coupling between operations and the
server that handles them -- \emph{this is a good thing}.  In any real
distributed system, we can't rely on having the same server handle multiple
sequential calls, if our abstraction makes this obvious, all the better.

A na\"{\i}ve implementation of our lookup function would force us to re-perform
a lookup for every request, which would likely have bad performance
implications.  This is easily solved by supplying a wrapper function that
provides caching.

\section{Conclusion}
Name resolution isn't just for domains anymore -- our RPC libraries should be
extended to support more dynamic, interesting forms of resoution.  The benefit
is simpler, more robust client code and better code reuse.

\bibliography{ref}
\bibliographystyle{abbrv}
\end{document}